\documentclass{article}
\usepackage{amsmath,amssymb}
\newcommand{\be}{\begin{equation}}
\newcommand{\ee}{\end{equation}}
\newcommand{\bea}{\begin{eqnarray}}
\newcommand{\eea}{\end{eqnarray}}
\newcommand{\bean}{\begin{eqnarray*}}
\newcommand{\eean}{\end{eqnarray*}}

\def\beq{\begin{equation}}

\def\eeq{\end{equation}}
\def\R{\mathcal{R}}

\relax

\def\d{\partial}

\def\a{\alpha'}

\begin{document}

\title{String-corrected dilatonic black holes in $d$ dimensions}

\author{Filipe Moura
\\
Centro de Matem\'atica da Universidade do Minho, \\Escola de Ci\^encias, Campus de Gualtar, \\4710-057 Braga, Portugal\\
\\
\texttt{fmoura@math.uminho.pt}
}
\maketitle

\begin{abstract}
We solve the dilaton field equation in the background of a spherically symmetric black hole in bosonic or heterotic string theory with curvature-squared corrections in arbitrary $d$ spacetime dimensions. We then apply this result to obtain a spherically symmetric black hole solution with dilatonic charge and curvature-squared corrections in bosonic or heterotic string theory compactified on a torus. For this black hole we obtain its free energy, entropy, temperature, specific heat and mass.
\end{abstract}

\section{Introduction}
Black holes have been over the last years in many ways the best object of predictions from string theory. Some of
such predictions are based on considering stringy effects, and how
they affect the classical black hole solutions and their
properties.

A frequently considered stringy effect is the result of
corrections in the inverse string tension ($\a$) in the form of
higher-derivative terms in the effective action. Curvature--squared
corrections to spherically symmetric $d-$dimensional black holes
in string theory were first discussed in \cite{cmp89}. This
article only addresses the effect of the $\a$ corrections; no
other string effects are considered. More recently, article
\cite{Giveon:2009da} considers other typical stringy effects,
namely string momentum and winding after compactification of a
fundamental string on an internal circle and $T$--dualization.

In this article we wish to study the effects on spherically symmetric black holes of
string compactification on a torus from 10 (or 26) to arbitrary
$d$ dimensions. In such compactification, one must pass to the
string to the Einstein frame, by a conformal transformation on the
original 10 (or 26) dimensions involving the dilaton field. Therefore we need to be in the presence of a dilaton field. We must then determine the solution to the dilaton in the background of a spherically symmetric black hole. We show that the dilaton vanishes classically and, therefore, one must really consider higher--curvature terms. This result had already been anticipated in \cite{cmp89}, where the authors take black holes directly in $d$ dimensions and just suggest, but do not fully consider, the effects of string compactification and of the presence of the dilaton.

The article is organized as follows: in section 2, we will solve
the dilaton field equation, in the background of a spherically
symmetric black hole in $d$ dimensions, in the presence of
curvature--squared corrections. Next, in section 3 we find out how
the presence of such dilaton actually changes the black hole in
bosonic or heterotic string theory, by considering such strings
compactified on a torus. Finally we derive some thermodynamical
properties of such black hole solution (free energy, entropy, temperature, specific heat and mass), which we compare to the equivalent results of the similar (nondilatonic) solution of \cite{cmp89}.

\section{The dilaton in the background of a $d-$dimensional black hole with $\R^2$ corrections}

The most general static, spherically symmetric metric in $d$ spacetime
dimensions can be written in spherical coordinates as
\be
d\,s^2=-f(r)\,d\,t^2 + g^{-1}(r)\,d\,r^2 +r^2\,d\,\Omega^2_{d-2}.
\label{metric1} \ee $f, g$ are arbitrary functions of the radius
$r;$ $d\,\Omega^2_{d-2}=\sum_{i=2}^{d-1} \frac{\prod_{j=2}^i
\sin^2 \theta_j}{\sin^2 \theta_i} d\,\theta_i^2$ is the element of
solid angle in the $(d-2)-$sphere. For pure Einstein-Hilbert
gravity in vacuum, the solution to the Einstein equations gives
\cite{Tangherlini:1963bw}
\be
f(r) = g(r) = 1 - \left(\frac{r_H}{r}\right)^{d-3},
\label{tangher} \ee $r_H$ being the horizon radius. This is the
$d-$dimensional extension of Schwarzschild's solution.

We are interested in extending this solution in the presence of a
dilaton, but considering string-theoretical $\a$ corrections. We
are focusing in particular in $\R^{\mu\nu\rho\sigma}
\R_{\mu\nu\rho\sigma}$ corrections, to first order in $\a$, which
are present in bosonic and heterotic string effective actions (but
not on type II superstring) \cite{Zwiebach:1985uq}. The effective
action we are thus considering, in the Einstein frame, is

\be \label{eef} \frac{1}{16 \pi G} \int \sqrt{-g} \left( \R -
\frac{4}{d-2} \left( \d^\mu \phi \right) \d_\mu \phi +
\mbox{e}^{\frac{4}{d-2} \phi} \frac{\lambda}{2}\
\R^{\mu\nu\rho\sigma} \R_{\mu\nu\rho\sigma} \right) \mbox{d}^dx .
\ee Here $\lambda = \frac{\a}{2}, \frac{\a}{4}$ and $0$, for
bosonic, heterotic and type II strings, respectively. We are only
considering gravitational terms: we can consistently settle all
fermions and gauge fields to zero for the moment. That is not the
case of the dilaton, as it can be seen from the field equations
(neglecting terms which are quadratic in $\phi$): \bea \nabla^2
\phi - \frac{\lambda}{4}\ \mbox{e}^{\frac{4}{2-d} \phi} \left(
\R_{\rho\sigma\lambda\tau} \R^{\rho\sigma\lambda\tau} \right) &=&
0, \label{bdfe} \\ \R_{\mu\nu} + \lambda\ \mbox{e}^{\frac{4}{2-d}
\phi} \left( \R_{\mu\rho\sigma\tau} {\R_{\nu}}^{\rho\sigma\tau} -
\frac{1}{2(d-2)} g_{\mu\nu} \R_{\rho\sigma\lambda\tau}
\R^{\rho\sigma\lambda\tau} \right) &=& 0. \label{bgfe} \eea From
(\ref{bdfe}), a constant-dilaton solution would imply the
vanishing of the source term $\R^{\mu\nu\rho\sigma}
\R_{\mu\nu\rho\sigma}:$ $\phi=0$ is a consistent solution only if
one takes $\lambda=0.$ This is the solution for $\phi$ we take at
this order. For the particular, spherically symmetric case we are
considering, we take (\ref{metric1}) with $f(r)=g(r)$ given by
(\ref{tangher}) as the $\lambda=0$ metric. We are interested in
computing the first $\a$ corrections to $\phi$ and $g_{\mu\nu},$
using (\ref{bdfe}) and (\ref{bgfe}) and always working
perturbatively in $\lambda,$ neglecting $\lambda^2$ and higher
order terms.

In order to avoid ghosts, the gravitational correction should in
principle be given by the Gauss-Bonnet combination
$\R^2_{GB}:=\R^{\mu\nu\rho\sigma} \R_{\mu\nu\rho\sigma} - 4
\R^{\mu\nu} \R_{\mu\nu} + \R^2$ but, since we are only interesting
in computing first-order perturbative in $\a$ corrections to a
classical solution, one can neglect the Ricci terms in the
low--energy effective action, which from (\ref{bgfe}) would only
contribute at a higher order in $\a;$ also, as a simple
computation shows, for the particular solution (\ref{tangher})
which we take as a background both terms are equivalent, since for this case one
has $4 \R^{\mu\nu} \R_{\mu\nu} \equiv \R^2$ and, therefore,
\be \label{r2f} \R^2_{GB} \equiv \R^{\mu\nu\rho\sigma}
\R_{\mu\nu\rho\sigma} = \frac{2 (d-2) (d-3)}{r^4} (1-f)^2 +\frac{2
(d-2)}{r^3} f'^2 +\left(f''\right)^2=(d-2)^2 (d-3) (d-1)
\frac{r_H^{2d-6}}{r^{2d-2}}. \ee

We get then $\nabla^\mu \nabla_\mu \phi(r) = \left(f
\phi'\right)'+\frac{d-2}{r} f \phi',$ from which $r^{d-2}
\nabla^\mu \nabla_\mu \phi(r) = \left( r^{d-2} f \phi' \right)'.$
We can take the $\lambda=0$ metric in (\ref{bdfe}) in order to
compute both $\nabla^\mu \nabla_\mu \phi(r),$ since $\phi$ is of
order $\lambda,$ and $\R^{\mu\nu\rho\sigma} \R_{\mu\nu\rho\sigma}$
(given in (\ref{r2f})), since this term is already multiplied by
$\lambda.$ Putting everything together, we write (\ref{bdfe}) as
\be
\left( \left(r^{d-2} -r^{d-3}_H r\right) \phi'\right)'= \lambda
\frac{(d-2)^2 (d-3) (d-1)}{4} \frac{r_H^{2d-6}}{r^d}, \ee which we
simply integrate to obtain
\be
\left(r^{d-2} -r^{d-3}_H r\right) \phi' = - \lambda \frac{(d-2)^2
(d-3)}{4} \frac{r_H^{2d-6}}{r^{d-1}} - (d-3) \Sigma.
\label{dilinha} \ee The integration constant $\Sigma,$ as will
become clear below, is the dilatonic charge. Integrating again,
and defining the incomplete Euler beta function as
$B(x;\,a,b)=\int_0^x t^{a-1}\,(1-t)^{b-1}\,dt,$ we find \bea
\phi(r) &=& -\frac{\Sigma}{r_H^{d-3}} \ln\left( 1 -
\left(\frac{r_H}{r}\right)^{d-3} \right) \nonumber \\
&-&\frac{\lambda}{r_H^2} \frac{(d-2)^2}{8}
\left[(d-3)\left(\frac{r_H}{r}\right)^2 +2 \frac{d-3}{d-1}
\left(\frac{r_H}{r}\right)^{d-1} - 2 B
\left(\left(\frac{r_H}{r}\right)^{d-3};\, \frac{2}{d-3}, 0
\right)\right]. \label{dil} \eea At asymptotic infinity this
solution is approximately given by
\be
\phi(r) \approx \frac{\Sigma}{r^{d-3}} + \frac{\Sigma r_H^{d-3}}{2
r^{2d-6}}+ \frac{\lambda}{8}(d-2)(d-3)\frac{r_H^{2d-6}}{r^{2d-4}},
\label{dildeser} \ee which is the asymptotic limit found in
\cite{bd86}. This solution depends on another parameter, the
dilatonic charge $\Sigma$, besides the black hole parameters $r_H$
and $\lambda,$ which could in principle be a sign for primary
hair. However, having a black hole solution means one only has a
coordinate (but not curvature) singularity at the horizon. From
the dilaton field equation (\ref{bdfe}), then, also $\phi(r)$ and
$\phi'(r)$ must be nonsingular at $r_H.$ From (\ref{dilinha}) we
see that, in order to avoid $\phi'$ becoming infinite at $r=r_H,$
one must choose an adequate value for $\Sigma$, given by
\be
\Sigma=-\frac{(d-2)^2}{4} \lambda r_H^{d-5}. \label{sigma} \ee Equation
(\ref{dil}) with $\Sigma$ given by (\ref{sigma}) is the solution
for the dilaton in the background of a spherically symmetric black
hole with $\R^{\mu\nu\rho\sigma} \R_{\mu\nu\rho\sigma}$
corrections in $d$ dimensions. This dilaton solution acts as
secondary hair, since it does not introduce any new physical
parameter besides the ones of the black hole. The parameter
$\frac{2}{d-3}$ is an integer for $d=4, 5;$ only for these values
of $d$ the function $B \left(\left(\frac{r_H}{r}\right)^{d-3};\,
\frac{2}{d-3}, 0 \right)$ can be written in terms of elementary
functions of calculus.\footnote{It is interesting to notice that
exactly the same argument can be used to show that, also for
toroidally compactified string theory, finite-horizon-area black
holes which are asymptotically flat only exist (in the
supergravity approximation, without $\a$ corrections) for $d=4,
5.$ In such approximation, one has
$g_{rr}=\left(1+\left(\frac{M}{r-r_H}\right)^{d-3}\right)^{\frac{2}{d-3}}.$
Solutions obtained by string theory compactifications always give
a positive integer as an exponent; therefore, in the same way
$\frac{2}{d-3}$ must be an integer. See \cite{Becker:2007zj},
chapter 11.} In particular, for $d=4$ our solution matches
perfectly the result of \cite{ms93}, as it should.

At the horizon, $\phi$ is indeed regular and given by
\footnote{The digamma function is given by
$\psi(z)=\Gamma'(z)/\Gamma(z),$ $\Gamma(z)$ being the usual
$\Gamma$ function. For positive $n,$ one defines
$\psi^{(n)}(z)=d^n\,\psi(z)/d\,z^n.$ This definition can be
extended for other values of $n$ by fractional calculus analytic
continuation. These are meromorphic functions of $z$ with no
branch cut discontinuities.

$\gamma$ is Euler's constant, defined by $\gamma=\lim_{n \to
\infty} \left(\sum_{k=1}^n \frac{1}{k} - \ln n \right),$ with
numerical value $\gamma \approx 0.577216.$}
\be
\phi\left(r_H\right)=-\frac{\lambda}{r_H^2} \frac{(d-2)^2}{8
(d-1)} \left(d^2-2d+2 (d-1) \left(\psi^{(0)}\left(\frac{2}{d-3}\right) + \gamma \right) -3\right). \label{firh}
\ee For $d=4,$ $\psi ^{(0)} \left(2\right)=1-\gamma;$ for $d=5,$
$\psi ^{(0)}\left(1\right)=-\gamma.$ Again, for higher values of
$d,$ $\phi\left(r_H\right)$ depends explicitly on $\gamma$ and
$\psi^{(0)}\left(\frac{2}{d-3}\right),$ but for $d=4,5$ this
dependence can always be eliminated. The same is true in general
for other expressions that we will meet later.

From (\ref{dilinha}) and (\ref{sigma}), the derivative of the
dilaton is given by $$\phi'\left(r\right)=\lambda \frac{(d-3)
(d-2)^2}{4} \frac{r_H^{d-3}}{r^d}
\frac{1-\left(\frac{r}{r_H}\right)^{d-1}}{1-\left(\frac{r}{r_H}\right)^{d-3}},$$
a strictly positive function for $r > r_H;$ we conclude that,
outside the horizon, $\phi$ grows between $\phi\left(r_H\right)$
given by (\ref{firh}) and 0, its value at infinity.

The article \cite{cmp89} determines the equivalent dilaton
solution in the string frame, where the field equations for the
dilaton are different than in the Einstein frame we are
considering. The final expression is relatively complicated; since
we don't need it here, we refer to the appendix of \cite{cmp89}.
The two solutions can be mapped by a transformation of the horizon
radius and are equivalent up to a shift by a constant value which
depends on $d.$ (The horizon radii are different in the two
frames; the relation between them can be seen explicitly in
\cite{Giveon:2009da}.) Because of such shift, the solution
(\ref{dil}) we present here, besides being more concise and
elegant, is the one which is normalized to vanish asymptotically,
according to (\ref{dildeser}).
\section{A black hole solution with $\R^2$ corrections for toroidal compactifications}

It would be interesting to obtain the $\a-$corrected black hole
solution coupled to the $\a-$corrected dilaton. In ref.
\cite{bd86} only approximations (at asymptotic infinity and close
to the horizon) are obtained. But in this article the authors
consider a primary-hair kind of dilaton, like (\ref{dildeser}),
but with $\Sigma$ as an independent parameter. The dilaton
solution of \cite{bd86} seems to be nonvanishing already at order
$\lambda=0,$ but the true physical solution (the only one which is
nonsingular at the horizon) is the one we have taken, with
$\Sigma$ given by (\ref{sigma}). Because $\Sigma$ depends on
$\lambda,$ this solution vanishes at order $\lambda=0.$ Having a
nonvanishing dilaton only at order $\lambda$ means that, when
solving the field equations, one can discard several terms
depending on $\phi$ in the perturbative expansion, which were not
discarded in \cite{bd86}.

The article \cite{cmp89} presents the $\lambda-$corrected metric,
for the system we are considering, in the Einstein frame, in a
system of coordinates such that the horizon radius $r_H$ is fixed
and has no $\a$ corrections. The result (the Callan--Myers--Perry
solution) is of the form (\ref{metric1}), with $f(r)=g(r) \equiv g_{CMP}(r),$ where
\be
\label{fr2} g_{CMP}(r) = \left(1-\left(\frac{r_H}{r}\right)^{d-3} \right) \left[1- \lambda
\frac{(d-3)(d-4)}{2} \frac{r^{d-5}_H}{r^{d-1}}
\frac{r^{d-1}-r_H^{d-1}}{r^{d-3} - r^{d-3}_H} \right]. \ee

This article \cite{cmp89} only considers (bosonic and heterotic)
string theory black hole solutions on arbitrary spacetime
dimensions in the presence of a dilaton. No other string effects
are considered. Recently, the article \cite{Giveon:2009da} has
considered black holes in any dimension formed by a fundamental
string compactified on an internal circle with any momentum $n$
and winding $w,$ both at leading order and with leading $\a$
corrections, by adding an additional coordinate to the solution of
\cite{cmp89}, boosting along this direction, reducing again to $d$
dimensions, $T$--dualizing (to change string momentum into
winding) and then boosting one other time to add momentum charge.
Einstein-Maxwell-dilaton black holes with $\R^2$ corrections in
any dimension have been considered in \cite{Chen:2009rv}. But no
solution considers the effects of  string compactification from 10
or 26 to $d$ dimensions.

String theories live in $d_s$ dimensions, with $d_s=26$ for
bosonic and $d_s=10$ for heterotic strings. When one talks about a
black hole in string theory in $d$ dimensions, the original
$d_s$--dimensional spacetime must have been compactified on some
$(d_s-d)$--dimensional manifold, with internal coordinates $y^m$
and internal metric $g_{mn}(y).$ When passing from the string to
the Einstein frame, one needs a transformation under which
\be
g_{\mu\nu} \rightarrow \exp \left( \frac{4}{d-2} \Phi \right)
g_{\mu\nu}, \,\, {\R_{\mu\nu}}^{\rho\sigma} \rightarrow
{\widetilde{\R}_{\mu\nu}}^{\ \ \ \rho\sigma} =
{\R_{\mu\nu}}^{\rho\sigma} -
{\delta_{\left[\mu\right.}}^{\left[\rho\right.} \nabla_{\left.\nu
\right]} \nabla^{\left.\sigma \right]} \Phi. \label{sf} \ee If one
takes this as a conformal transformation of the entire
$d_s-$dimensional metric (rather than just on the $d-$dimensional
black hole part, as it was done in \cite{cmp89} to obtain
(\ref{fr2})), it involves the total dilaton field $\Phi,$
including the Kaluza-Klein part depending on the internal
coordinates $y^m$ (rather than just the $d-$dimensional part
$\phi$ as we have been considering). This way the size of the
compact space becomes spatially varying, being governed by a
function $h.$ The total metric is then of the form
\be
d\,s^2=-f(r)\,d\,t^2 + g^{-1}(r)\,d\,r^2+r^2\,d\,\Omega^2_{d-2} +
h \, g_{mn}(y)\, d\,y^m\,d\,y^n. \label{metric2} \ee Taking the
field equations for the whole spacetime, the compact space and the
black hole are no longer decoupled, due to the term $g_{\mu\nu}
\R_{\rho\sigma\lambda\tau} \R^{\rho\sigma\lambda\tau}$ in
(\ref{bgfe}). In order to avoid this problem, we take the internal
space to be a flat torus, with vanishing internal curvature to
leading order. If this is the case, the function $h$ can be shown
to depend only on the $d-$dimensional part of the dilaton $\phi,$
i.e. $h=h(\phi).$ The solution (\ref{metric2}) to (\ref{bgfe}) is
then \cite{cmp89} \bea h(\phi)&=&\left(1-\frac{2}{d_s-2} \phi
\right)^2, \\ f(r) &=& g(r) + 4 \left( 1 -
\left(\frac{r_H}{r}\right)^{d-3} \right)
\frac{d_s-d}{\left(d_s-2\right)^2} \left(\phi - r \phi'\right),
\label{fnew} \eea with $g(r)$ being equal to $g_{CMP}(r)$ given by
(\ref{fr2}). Explicitly, using the dilaton solution (\ref{dil})
for $\phi$, \bea f(r) &=& \left( 1 -
\left(\frac{r_H}{r}\right)^{d-3} \right) \left( 1 -
\frac{(d-3)(d-4)}{2} \ \frac{\lambda}{r_H^2}
\left(\frac{r_H}{r}\right)^{d-3}\ \frac{1 - \left(
\frac{r_H}{r}\right)^{d-1}}{1 -
\left(\frac{r_H}{r}\right)^{d-3}}\right. \nonumber\\ &+& (d-2)^2\
\frac{d_s-d}{\left(d_s-2\right)^2}\ \frac{\lambda}{r_H^2} \left[
\ln\left( 1 - \left(\frac{r_H}{r}\right)^{d-3} \right) +
B\left(\left(\frac{r_H}{r}\right)^{d-3};\,\frac{2}{d-3}, 0
\right)\right. \nonumber
\\ &-& \left. \left.
\frac{d-3}{2}\ \left( \frac{r_H}{r}\right)^2 - \frac{d-3}{d-1}
\left(\frac{r_H}{r}\right)^{d-1} -(d-3)
\left(\frac{r_H}{r}\right)^{d-3}\ \frac{1 - \left(
\frac{r_H}{r}\right)^{d-1}}{1 - \left(\frac{r_H}{r}\right)^{d-3}}
\right] \right). \eea As one sees from (\ref{fnew}), when $d=d_s$
one has $f(r)=g(r).$ This is to be expected: in this case there is
no compactification, and (\ref{metric2}) reduces to
(\ref{metric1}), with $f(r)=g(r)=g_{CMP}(r)$ given by (\ref{fr2}).

\section{Thermodynamical properties}
In this section, we compute several thermodynamical quantities for
the black hole solution we have just found. In each case we
compare the result to the corresponding one of the solution
(\ref{fr2}) obtained in \cite{cmp89}, since the parameters are the
same. This way we can evaluate the physical effects introduced by
the toroidal compactification.

The free energy of the black hole solution (\ref{fnew}) is
obtained from the euclideanized Einstein-frame action (\ref{eef}),
to which one adds a surface term consisting of an integral (on the
boundary) of the trace of the second fundamental form, subtracted
by the same trace for the boundary embedded on flat space, to
render the total surface contribution finite. This surface term
also includes contributions for the higher-derivative terms, but
these contributions do not affect this calculation. Also, there
exists a choice of fields such that all the terms in the euclidean
action directly involving the dilaton, the Kaluza-Klein scalar and
the $(d_S-d)-$torus metric are of order $\lambda^2$ and can,
therefore, be neglected. This means in particular that the result
for the free energy for our solution is the same that for the
solution (\ref{fr2}), whose calculation is described in
\cite{cmp89}; the result is ($\Omega_{d-2}=2
\pi^{\frac{d-1}{2}}/\Gamma\left(\frac{d-1}{2}\right)$ being the
area of the unit $(d-2)-$sphere) \cite{Giveon:2009da} \be
\label{pmef} F=\left(1-\frac{d(d-3)}{2} \frac{\lambda}{r_H^2}
\right) \frac{\Omega_{d-2}}{16 \pi G} r^{d-3}_H. \ee

The entropy of this black hole solution can be obtained by Wald's
formula \cite{w93} $$S=-2 \pi \int_H \frac{\partial
\mathcal{L}}{\partial R^{\mu \nu \rho \sigma}}
\varepsilon^{\mu\nu} \varepsilon^{\rho\sigma} \, \sqrt{h} \,
d\,\Omega_{d-2},$$ since from (\ref{eef}) we are dealing with a
lagrangian $\mathcal{L}$ with higher derivatives. $H$ is the black
hole horizon, with area $A_H=r_H^{d-2} \Omega_{d-2}$ and metric
$h_{ij}$ induced by the spacetime metric $g_{\mu\nu}.$ For the
metric (\ref{metric2}), the nonzero components of the binormal
$\varepsilon^{\mu\nu}$ to $H$ are
$\varepsilon^{tr}=-\varepsilon^{rt}=-\sqrt{\frac{g}{f}}.$ From
(\ref{eef}) one also needs $$8 \pi G\frac{\partial
\mathcal{L}}{\partial R^{\mu \nu \rho
\sigma}}=\frac{1}{4}\left(g_{\mu\rho}g_{\sigma\nu}-g_{\mu\sigma}g_{\rho\nu}\right)+
\mbox{e}^{\frac{4}{d-2} \phi} \frac{\lambda}{2} R_{\mu \nu \rho
\sigma}.$$This way, taking only nonzero components, one gets from
(\ref{metric2}) $$8 \pi G \frac{\partial \mathcal{L}}{\partial
R^{\mu \nu \rho \sigma}} \varepsilon^{\mu\nu}
\varepsilon^{\rho\sigma} = 4 \times 8 \pi G \frac{\partial
\mathcal{L}}{\partial R^{trtr}} \varepsilon^{tr} \varepsilon^{tr}
=\left(-\frac{f}{g}+\mbox{e}^{\frac{4}{d-2} \phi} \lambda
f''\right)\frac{g}{f}.$$ At order $\lambda=0,$ $\phi=0,$ $f=g$ and
$f''=-\frac{1}{r_H^2} (d-3) (d-2).$ Therefore
\be
S=\frac{1}{4 G} \int_H \left( 1 +\frac{\lambda}{r_H^2} (d-3)
(d-2)\right) \, \sqrt{h} \, d\,\Omega_{d-2} =\frac{A_H}{4 G}
\left(1+ (d-3) (d-2) \frac{\lambda}{r_H^2}\right). \label{pmes}
\ee Because the $\lambda$--correction to the entropy depends only
on the $\lambda=0$ part of the metric, it is no surprise that this
same result was obtained (by a different process, though, and for
a metric (\ref{fr2}) with a different $\a$ correction) in
\cite{cmp89}.

In order to compute the black hole temperature, one first
Wick--rotates the metric (\ref{metric2}) to Euclidean time $t = i
\tau$. The resulting manifold has no conical singularities as long
as $\tau$ is a periodic variable, with a period $\beta$ related to
the black hole temperature as $T = \frac{1}{\beta}$. The precise
smoothness condition is $2 \pi = \lim_{r \to r_H}
\frac{\beta}{g^{-\frac{1}{2}} \left(r\right)} \frac{d
f^{\frac{1}{2}}\left(r\right)}{d r},$ from which one gets $T=
\lim_{r\rightarrow r_H}\frac{\sqrt{g}}{2\pi}
\frac{d\,\sqrt{f}}{d\,r}.$ In our particular case,
\bea
\label{pmet} T &=& \frac{d-3}{4 r_H
\pi}\left[1+\frac{\lambda}{r_H^2} \delta T(d,d_s) \right],
\nonumber \\ \delta T(d,d_s)&=&  \frac{1}{4 (d-1)
\left(d_s-2\right)^2} \left[ 3 d^5-(3 d_s +18)
d^4+\left(-2 d_s^2 +26 d_s +27\right) d^3 \right. \nonumber \\
&+& \left(12 d_s^2-83 d_s+28\right) d^2
- 2 \left(9 d_s^2-46 d_s+38\right) d  +4 \left(2 d_s^2-7 d_s+8\right)
\nonumber \\
&+& \left. 2 (d-2)^2 (d-1) (d-d_s) \left(\psi^{(0)}\left(\frac{2}{d-3}\right)+ \gamma \right)
\right]. \eea
We have checked that
$\delta T$ and, therefore, the correction term to the temperature,
are always negative: $\a$ corrections decrease the temperature for
every relevant values of $d$ and $d_s.$ If one takes the approximate expression
(\ref{pmet}) as exact, one may even get $T<0$ to first order in
$\lambda$ for some values of $d, d_s$ and $\lambda/r_H^2.$ From
our evaluation of $\delta T,$ we concluded that the approximate
expression for $T$ given by (\ref{pmet}) is positive as long as
$\a<0.148148 r_H^2$ (for $d_s=10$) or $\a<0.00727273 r_H^2$ (for
$d_s=26$). But (\ref{pmet}) is only a first--order perturbative
approximation; a complete analysis would require a full knowledge
of $T$ to all orders. Nonetheless, the leading string correction
being negative suggests that the temperature may reach a maximum,
for each particular given value of $d, d_s,$ approximately for
$r_H= \sqrt{-3 \delta T(d,d_s) \lambda}$ (again taking
(\ref{pmet}) as an exact expression, a good approximation if the
higher--order $\a$ corrections are much smaller than the
first--order one we are considering, something that should be true
at least for large black holes). For all possible values of $d$
and $d_s,$ we evaluated $T=\frac{d-3}{6 r_H \pi}$ (which is what
one obtains after replacing $r_H= \sqrt{-3 \delta T(d,d_s)
\lambda}$ in (\ref{pmet})). For $d_s=10$ we obtained a maximum
$T_{\mathrm{max}}=\frac{0.082}{\sqrt{\a}}$ for $d=10$, while for
$d_s=26$ we obtained a maximum
$T_{\mathrm{max}}=\frac{0.071}{\sqrt{\a}}$ for $d=4.$ Like the
ones corresponding to the solution (\ref{fr2}) determined in
\cite{cmp89}, these values are smaller than the critical Hagedorn
temperatures, obtained from the free string spectrum, and given by
$T_{\mathrm{crit}}=\frac{0.16}{\sqrt{\a}}$ (for the heterotic
string, with $d_s=10$) and
$T_{\mathrm{crit}}=\frac{0.08}{\sqrt{\a}}$ (for the bosonic
string, with $d_s=26$).

It is interesting to compare the value $\delta T(d,d_s)$ we
obtained with the corresponding one for the noncompactified solution
(\ref{fr2}) from \cite{cmp89}. For this solution, the temperature
is given by
\be
T_{CMP} = \frac{d-3}{4 r_H \pi}\left[1-\frac{\lambda}{r_H^2}
\frac{(d-1)(d-4)}{2} \right]. \ee
We have checked that $\delta T(d,d_s) < -\frac{(d-1)(d-4)}{2},$ i.e. the decrease in $T$ due to $\a$ corrections is larger for (\ref{fnew}) than for (\ref{fr2}), for
every relevant values of $d$ and $d_s.$  The only exception is precisely when $d=d_s,$ when $\delta T(d,d_s) \equiv -\frac{(d-1)(d-4)}{2},$ for the reasons we have already mentioned.

The specific heat is given by $C= T
\frac{\partial\,S}{\partial\,T} =T
\frac{\frac{d\,S}{d\,r_H}}{\frac{d\,T}{d\,r_H}}.$ In our case, one
is left with \bea C&=& -(d-2)\frac{A_H}{4G}
\left[1+\frac{\lambda}{r_H^2} \delta C(d,d_s)\right],\nonumber \\
\delta C(d,d_s) &=& - \frac{d-2}{2 (d-1) (d_s-2)^2} \left[62 d_s- 16d_s^2-64 +3 d^4 -3 (d_s+4) d^3 -(2 d_s (2 d_s
-14) +5) d^2\right.\nonumber \\
&+& \left. (d_s (20 d_s -91) +82) d + 2 (d-1)
(d-2) (d-d_s) \left(\psi^{(0)}\left(\frac{2}{d-3}\right) + \gamma \right)\right].
\label{pmec} \eea We checked that $\delta C(d,d_s)$ is always
positive for every relevant value of $d$ and $d_s,$ which means
$\a$--corrected black holes keep being thermodynamically unstable.

For the noncompactified solution (\ref{fr2}), the specific heat is given by
\be
C_{CMP} =  -(d-2)\frac{A_H}{4G}
\left[1+2(d-4)(d-2) \frac{\lambda}{r_H^2}\right]. \ee
It is also interesting to compare the value $\delta C(d,d_s)$ we
obtained with the corresponding one for the noncompactified solution
(\ref{fr2}). We checked that $\delta C(d,d_s) > 2(d-4)(d-2),$ for
every relevant value of $d$ and $d_s$ except when $d=d_s.$ This means the $\a$ correction is bigger, i.e. $C$ becomes more negative with (\ref{fnew}) than with (\ref{fr2}).

The black hole inertial mass matches the result of solution
(\ref{fr2}) of \cite{cmp89}:
\be
M_I = M_{CMP}=\frac{\left( d-2
\right) \Omega_{d-2}}{16 \pi G} \lim_{r \to \infty} r^{d-3} \Big(
1 - g \left(r\right) \Big) 
= \left( 1 + \frac{(d-3)(d-4)}{2}\
\frac{\lambda}{r_H^2} \right) \frac{\left(d-2\right)
\Omega_{d-2}}{16 \pi G} r^{d-3}_H. \label{mcmp}
\ee
Since $g(r) \neq f(r),$ one
expects the black hole inertial and gravitational masses to be
different. This situation is usual when one is dealing with
compactifications and originates from the integration of
Kaluza-Klein modes in the full $d_s-$dimensional action, resulting
in a $d-$dimensional action with nondiagonal kinetic terms.
Indeed, from (\ref{dildeser}) and (\ref{fnew}), one gets
\be
M_G = \frac{\left( d-2 \right) \Omega_{d-2}}{16 \pi G} \lim_{r \to
\infty} r^{d-3} \Big( 1 - f \left(r\right) \Big) =M_I +
\frac{d_s-d}{\left(d_s-2\right)^2}\frac{\left( d-2 \right)^4
\Omega_{d-2}}{16 \pi G} \frac{\lambda}{r_H^2} r^{d-3}_H.
\ee

The actual physical mass is obtained by the relation $M=ST+F.$ From (\ref{pmef}), (\ref{pmes}), (\ref{pmet}),
\bea
M&=&\left[1+\frac{\lambda}{r_H^2} \delta M(d,d_s)\right] \frac{\left(d-2\right) \Omega_{d-2}}{16 \pi G} r^{d-3}_H,  \nonumber\\
\delta M(d,d_s)&=& \left[ 3 d^4-3 (d_s+4) d^3+(2 d_s (d_s +2) +19) d^2 \right.+(d_s (-10 d_s +29)-38) d+2 d_s (4 d_s -17) \nonumber \\
&+& \left.2 (d-1) (d-2) (d-d_s) \left(\psi^{(0)}\left(\frac{2}{d-3}\right) + \gamma \right)+32 \right] \frac{(d-3)}{4 (d-1) (d_s-2)^2}. \label{pmem}
\eea
The sign of $\delta M(d,d_s)$ depends on its parameters. For $d=4$ and $d=5, d_s=10$ it is negative, i.e. $M$ decreases with $\a$ corrections; for $d=5, d_s=26$ and $d>5$ it is positive, which means $\a$ corrections increase $M.$ It is important to verify that, taking (\ref{pmem}) as an exact expression, one does not get a negative mass to first order in $\lambda,$ i.e. when $\delta M(d,d_s)$ is negative. We verified that for such cases, exactly like we did with the temperature. The limits are much less restrictive this time: $M$ given by (\ref{pmem}) is positive as long as $\a<8.82759\,r_H^2$ (for $d_s=10$) or $\a<10.8339\,r_H^2$ (for $d_s=26$). Again, a complete analysis would require a full knowledge of $M$ to all orders.

We also compared the value $\delta M(d,d_s)$ we
obtained with the corresponding one for the noncompactified solution
(\ref{fr2}) from \cite{cmp89} given by (\ref{mcmp}). We checked that $\delta M(d,d_s) < \frac{(d-3)(d-4)}{2},$ i.e. the increase in $M$ due to $\a$ corrections is smaller for (\ref{fnew}) than for (\ref{fr2}), for
every relevant value of $d$ and $d_s,$ except when $d=d_s.$

One can invert (\ref{pmem}) to get the horizon radius as a function of the mass, obtaining
\bea
r_H&=&\frac{8^{\frac{1}{d-3}}}{\sqrt{\pi}} \left(\frac{GM \Gamma\left(\frac{d-1}{2}\right)}{d-2}\right)^{\frac{1}{d-3}} \left[1-\frac{4^{\frac{d}{3-d}} \pi}{(d-1) (d_s-2)^2} \left(\frac{d-2}{GM \Gamma \left(\frac{d-1}{2}\right)}\right)^{\frac{2}{d-3}} \lambda \right. \nonumber \\
&&\left[3 d^4\right. -3 (d_s+4) d^3 + (2 d_s (d_s +2) +19) d^2+(d_s (-10 d_s +29) -38) d \nonumber \\
&+& \left. \left. 2 d_s (4 d_s -17) + 2 (d-1) (d-2) (d-d_s) \left(\psi^{(0)}\left(\frac{2}{d-3}\right) + \gamma \right)+32 \right] \right].
\label{rhm}
\eea
This expression must be interpreted with care. In (\ref{pmem}) we obtained the leading perturbative correction to $M$ as a function of $r_H,$ but only if we knew the full expression $M(r_H),$ including all the string corrections, could we eventually invert it, and obtain an expression for $r_H$ as a function of the full physical string--corrected mass (and not just the classical mass, given by setting $\lambda=0).$ Equation (\ref{rhm}) represents the leading term in a series, but it does not represent by itself a string correction to $r_H.$ This is because (\ref{rhm}) hides the fact that $M$ itself has string corrections. If one considers those string corrections on $M,$ they should be such that they would eventually be cancelled, at every order in $\lambda,$ when taken all together. Indeed, by assumption, $r_H$ receives no $\a$ corrections \cite{cmp89} - it is the only free parameter of the solution and, together with $\lambda, d, d_s$ determines all the physical quantities.

It is useful to express the thermodynamical quantities we have been computing in terms of the physical mass, by replacing (\ref{rhm}) in (\ref{pmef}), (\ref{pmes}), (\ref{pmet}) and (\ref{pmec}). The temperature is expressed as
\bea
T&=&\frac{2^{\frac{3-2 d}{d-3}} (d-3)}{\sqrt{\pi }} \left(\frac{d-2}{GM \Gamma \left(\frac{d-1}{2}\right)}\right)^{\frac{1}{d-3}} \left[1+\frac{2^{-\frac{2 d}{d-3}} (d-2)}{(d_s-2)^2} \pi \left(\frac{d-2}{GM \Gamma \left(\frac{d-1}{2}\right)}\right)^{\frac{2}{d-3}} \lambda
\left[8 d_s^2-31 d_s+ 32 \right. \right. \nonumber \\
&+& \left. \left. 3 d^3 - (3 d_s+6) d^2 - \left(2 d_s^2-14 d_s +9\right) d+
+  2 (d-1) (d-d_s) \left(\psi^{(0)}\left(\frac{2}{d-3}\right) + \gamma \right) \right] \right], \label{tm}
\eea
while the free energy, entropy and specific heat are given by
\bea
F&=&\frac{M}{d-2} \left[1-\frac{4^{\frac{d}{3-d}} (d-3) (d-2) \pi}{(d-1) (d_s-2)^2} \left(\frac{d-2}{GM \Gamma
\left(\frac{d-1}{2}\right)}\right)^{\frac{2}{d-3}} \lambda \left[2 (d-1) (d-d_s) \left(\psi^{(0)}\left(\frac{2}{d-3}\right) + \gamma \right) \right. \right. \nonumber \\
&+& \left. \left. 3 d^3-(3 d_s+6) d^2 +\left(4 d_s^2-10 d_s +15\right) d+d_s (-4 d_s +17)-16\right] \right];
\eea
\bea
S&=&2^{\frac{2 d-3}{d-3}} \frac{M}{d-2} \sqrt{\pi} \left(\frac{GM \Gamma \left(\frac{d-1}{2}\right)}{d-2}\right)^{\frac{1}{d-3}}
\left[1- \frac{4^{\frac{d}{3-d}} (d-2)^2 \pi \lambda}{(d-1)
   (d_s-2)^2} \left(\frac{d-2}{GM \Gamma \left(\frac{d-1}{2}\right)}\right)^{\frac{2}{d-3}} \right. \nonumber \\
&\times& \left[3 d^3-(3 d_s +6) d^2-(2 d_s (d_s -7) +9) d-7 d_s+2 d_s^2 +8 \right. \nonumber \\
&+& \left. \left. 2 (d-1) (d-d_s) \left(\psi^{(0)}\left(\frac{2}{d-3}\right) + \gamma \right) \right] \right];\\
C&=& -2^{\frac{2d-3}{d-3}} M \sqrt{\pi } \left(\frac{GM \Gamma \left(\frac{d-1}{2}\right)}{d-2}\right)^{\frac{1}{d-3}}
   \left[1-\frac{3 (d-2) 4^{\frac{d}{3-d}} \pi}{(d-1) (d_s-2)^2} \left(\frac{d-2}{GM \Gamma
   \left(\frac{d-1}{2}\right)}\right)^{\frac{2}{d-3}} \lambda \left[3 d^4 \right. \right. \nonumber \\
   &-&3 (d_s+4) d^3+(-2 d_s
   (d_s-10) +3) d^2+(d_s (10 d_s -51)+42) d \nonumber \\
   &+& \left. \left. 30 d_s-8 d_s^2 +2 (d-1) (d-2) (d-d_s) \left(\psi^{(0)}\left(\frac{2}{d-3}\right) + \gamma \right)-32\right]\right].
\eea
These variables can also be expressed in terms of the temperature: by inverting (\ref{pmet}) instead of (\ref{pmem}), one obtains an expression analogous to (\ref{rhm}) which, when replaced in (\ref{pmef}), (\ref{pmes}), and (\ref{pmec}), gives
\bea
F&=& \frac{2^{3-2d} \pi^{\frac{3-d}{2}}}{G \Gamma \left(\frac{d-1}{2}\right)} \left(\frac{d-3}{T}\right)^{d-3} \left[1+ \frac{4 (d-2)^2 \pi^2 \lambda T^2}{(d-3) (d-1) (d_s-2)^2} \left[2 (d-1) (d-d_s) \left(\psi^{(0)}\left(\frac{2}{d-3}\right) + \gamma \right) \right. \right. \nonumber \\
&+& \left. \left. \left(3 d^3-(3 d_s+6) d^2-(2 d_s (d_s -7) +9) d-7 d_s+2 d_s^2+8\right) \right] \right]; \\
S&=& \frac{2^{3-2 d} \pi^{\frac{3-d}{2}}}{G \Gamma\left(\frac{d-1}{2}\right)} \left(\frac{d-3}{T}\right)^{d-2} \left[1+\frac{4 (d-2)^2 \pi^2  \lambda T^2}{(d-3)^2 (d-1) (d_s-2)^2} \left[3 d^4-3 (d_s+4) d^3 \right. \right. \nonumber \\
&+& (-2 d_s (d_s -10) +3) d^2+(d_s (12 d_s -59) +50) d \nonumber \\
&-& 2\left. \left. (d_s (5 d_s -19)+20)+2 (d-1) (d-2) (d-d_s) \left(\psi^{(0)}\left(\frac{2}{d-3}\right) + \gamma \right)\right] \right]; \\
C&=& -\frac{2^{3-2 d} (d-2) \pi^{\frac{3-d}{2}}}{G\Gamma \left(\frac{d-1}{2}\right)} \left(\frac{d-3}{T}\right)^{d-2} \left[1+\frac{4 (d-2)(d-4) \pi ^2 \lambda T^2}{(d-3)^2 (d-1) (d_s-2)^2} \left[3 d^4-3 (d_s+4) d^3 \right. \right. \nonumber \\
&+& (-2 d_s (d_s -10) +3) d^2+(d_s (12 d_s -59) +50) d \nonumber \\
&-& 2\left. \left. (d_s (5 d_s -19)+20)+2 (d-1) (d-2) (d-d_s) \left(\psi^{(0)}\left(\frac{2}{d-3}\right) + \gamma \right)\right] \right]. \label{ct}
\eea
For all these expressions (\ref{tm})--(\ref{ct}), the same warning we made for (\ref{rhm}) applies: they are exact just for $\lambda=0$ (involving just the classical quantities, without any $\a$ corrections). Beyond the classical limit, these expressions just give an indication of the first--order (in terms of mass or temperature) terms of unknown functions, whose full expressions could only be determined if we knew all these quantities to all orders in $\a.$ The true first--order corrections in $\a$ are those given in equations (\ref{pmef})--(\ref{pmem}), in terms of $\lambda/r_H^2,$ whose signs and magnitudes we analyzed.

\section{Conclusions}

In this work, we derived the spherically symmetric solution to a dilaton in the presence of a black hole in string theory with curvature-squared corrections in $d$ spacetime dimensions. We then obtained a spherically symmetric black hole solution with dilatonic charge and curvature-squared corrections from compactified string theory in $d$ dimensions, and we computed its free energy, entropy, temperature, specific heat and mass. We compared the magnitude of the $\a$ corrections to these quantities to the ones corresponding to the noncompactified solution (\ref{fr2}) from \cite{cmp89}, in order to estimate the effects of string compactification. Free energy is decreased and entropy is increased with $\a$ corrections; the magnitude of the corrections is the same as in the solution (\ref{fr2}). Also like in (\ref{fr2}), the temperature decreases and the specific heat becomes more negative, but in our case the effects of the $\a$ corrections are strengthened. The $\a$ corrections to the mass, on the contrary, are weakened in comparison to (\ref{fr2}) (whose value for the mass is always increased), and for a few values of $d$ they even mean a decrease in $M.$

In a future work we plan to study some other features of this black hole, like its stability and scattering of gravitational waves.

\paragraph{Acknowledgements}
\noindent
This work has been supported by CMAT - U.Minho through the FCT Pluriannual Funding Program, and by FCT and FEDER through project CERN/FP/116377/2010. The author thanks Miguel Paulos for useful discussions.

\end{document}